\documentclass[aps,prl,column,showpacs,superscriptaddress,preprint]{revtex4-1}
\usepackage{graphicx}
\usepackage{natbib}
\usepackage{amsmath}
\usepackage{amssymb}
\usepackage{bm}
\usepackage{color}
\usepackage{float}

\begin{document}

\title{Unraveling photoinduced spin dynamics in topological insulator Bi$_2$Se$_3$}

\author{M. C. Wang} 
\affiliation{The Peac Institute of Multiscale Sciences, Chengdu, Sichuan 610031, P. R. China}
\affiliation{Key Laboratory of Advanced Technologies of Materials, Ministry of Education, Southwest Jiaotong University, Chengdu, Sichuan 610031, P. R. China
}
\author{S. Qiao} 
\affiliation{State Key Laboratory of Functional Materials for Informatics, Shanghai Institute of Microsystem and Information Technology, Chinese Academy of Sciences, Shanghai, Shanghai 200050, P. R. China}
\affiliation{School of Physical Science and Technology, ShanghaiTech University, Shanghai 200031, P. R. China}
\author{Z. Jiang}
\affiliation{School of Physics, Georgia Institute of Technology, Atlanta, Georgia 30332, USA}
\author{S. N. Luo} 
\affiliation{The Peac Institute of Multiscale Sciences, Chengdu, Sichuan 610031, P. R. China}
\affiliation{Key Laboratory of Advanced Technologies of Materials, Ministry of Education, Southwest Jiaotong University, Chengdu, Sichuan 610031, P. R. China
}
\author{J. Qi}
\email{jqi@pims.ac.cn} 
\affiliation{The Peac Institute of Multiscale Sciences, Chengdu, Sichuan 610031, P. R. China}
\affiliation{Key Laboratory of Advanced Technologies of Materials, Ministry of Education, Southwest Jiaotong University, Chengdu, Sichuan 610031, P. R. China
}

\begin{abstract}
We report on time-resolved ultrafast optical spectroscopy study of the topological insulator (TI) Bi$_2$Se$_3$. We unravel that a net spin polarization can not only be generated using circularly polarized light via interband transitions between topological surface states (SSs), but also via transitions between SSs and bulk states. Our experiment demonstrates that tuning photon energy or temperature can essentially allow for photoexcitation of spin-polarized electrons to unoccupied topological SSs with two distinct spin relaxation times ($\sim$25 fs and $\sim$300 fs), depending on the coupling between SSs and bulk states. The intrinsic mechanism leading to such distinctive spin dynamics is the scattering in SSs and bulk states which is dominated by $E_g^2$ and $A_{1g}^1$ phonon modes, respectively. These findings are suggestive of novel ways to manipulate the photoinduced coherent spins in TIs.
\end{abstract}
\maketitle

Topological insulators (TIs), as a new quantum phase of matter, are characterized by an unusual electronic structure exhibiting both insulating bulk and robust metallic surface states (SSs) \cite{Hasan_RMP_2010, Qi_RMP_2011}. This unique electronic structure combining external light excitation on TIs leads to many exotic physical phenomena \cite{Tse_PRL_2010, Aguilar_PRL_2012, Inoue_PRL_2010, Lindner_NatPhys_2011, Wang_Science_2013, McIver_NatNano_2011, Lu_PRB_2010, Hosur_PRB_2011, Junck_PRB_2013, Kastl_NatComm_2015}, which hold TIs a great promise for opto-spintronics and ultrafast spintronics applications \cite{Pesin_NatMat_2012}. Therefore, it becomes crucial to study the out-of-equilibrium properties of TIs under photo-excitation. Among them, the charge and spin dynamics have attracted a lot of recent attention, and are explored effectively using the time- and angle-resolved photoemission spectroscopy (Tr-ARPES) \cite{Sobota_PRL_2012, Wang_PRL_2012, Hajlaoui_NL_2012, Crepaldi_PRB_2012, Sobota_PRL_2013, Sobota_JESR_2014, Sobota_PRL_2014, Cacho_PRL_2015} and time-resolved optical spectroscopy \cite{Qi_APL_2010, Kumar_PRB_2011, Hsieh_PRL_2011, Chen_APL_2012, Glinka_APL_2013, Luo_NL_2013, Lai_APL_2014, Cheng_APL_2014, Sim_PRB_2014, Aguilar_APL_2015, Onishi_PRB_2015}. Investigation of the non-equilibrium charge dynamics enables deep understanding of the momentum scattering, which is a fundamental process determining the electronic transport in TIs \cite{Butch_PRB_2010}. On the other hand, comprehensive knowledge of the spin dynamics, including coherent spin generation and relaxation, is vital for actively manipulating spins in spintronics \cite{Zutic_RMP_2004}. Most of previous time-resolved works focus on the charge dynamics in TIs, and have revealed that the electron-phonon (e-p) coupling plays a key role in momentum scattering of the non-equilibrium carriers. In contrast, very few works pay attention on the photoinduced coherent spin dynamics in TIs \cite{Hsieh_PRL_2011, Boschini_arXiv_2015, Barriga_arXiv_2015}. Specifically, several key questions are still open regarding the dynamical response of the spin properties to the incident light: (1) Can a surface net spin polarization be generated via interband transitions from bulk states to SSs using circularly polarized light? (2) Does the spin dynamics in topological SSs and bulk states behave in a similar manner? Or does the same mechanism dictate the spin dynamics in topological SSs and bulk states?

In this Letter, we employ time-resolved transient reflectivity and Kerr rotation measurements to investigate the photo-excited charge and coherent spin dynamics in the prototypical TI Bi$_2$Se$_3$. Two types of helicity-dependent photoinduced net spin polarization have been unveiled, and can be attributed to the interband transitions between SSs and bulks and between SSs and SSs, respectively. We show that two coherent spin dynamics with distinct spin relaxation times can be selectively excited by tuning the photon energy $h\nu$ or temperature $T$. We reveal that the e-p scattering in SSs and bulk states dominated by different phonon modes gives rise to the distinct spin relaxation.   

Time-resolved transient reflectivity change $\Delta R(t)/R$ and time-resolved Kerr rotation (TRKR) $\Delta\theta_K(t)$ were measured based on a pump-probe scheme using a Ti:sapphire laser with a time resolution of $\sim$35 fs. Experiments were performed on high quality Bi$_2$Se$_3$ samples with carrier density $n\simeq 4\times10^{18}$ cm$^{-3}$. The Fermi level $E_f$ resides inside the bulk band-gap, and is $\sim$0.1 eV above the Dirac point. Detailed description of the materials and the measurement setup can be found in Supplemental Material \cite{Suppl}. 

\begin{figure}[h]
\includegraphics[width=12cm]{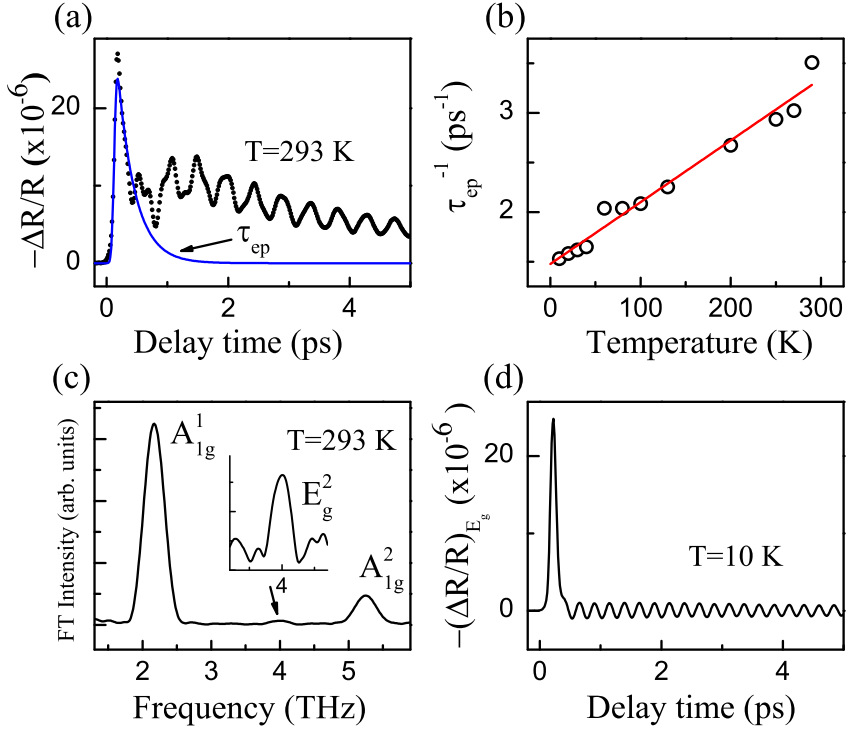}
\caption{\label{fig:deltaR} (a) Typical $\Delta R/R$ measurement on Bi$_2$Se$_3$ at 293 K with $h\nu\sim$1.55 eV. The blue line is an exponential fit to the initial fast decay. (b) Electron-phonon scattering rate $\tau_{ep}^{-1}$ as a function of temperature. The red line is a linear fit. (c) Fourier transform (FT) spectra of the oscillatory behaviour. (d) Typical $(\Delta R/R)_{E_g}$ signal associated with the $E_g$ phonon modes at 10 K.}
\end{figure}

Figure \ref{fig:deltaR}(a) shows a typical transient reflectivity $\Delta R/R$ signal, where an initial fast-decaying component is followed by slow relaxation processes superimposed with an oscillatory behavior. An exponential fit to the initial fast decay (blue line) exhibits a time constant of $\tau_{ep}\simeq$ 300 fs, consistent with the cooling time of electrons in the bulk bands for electronic temperature $T_e\gtrsim$600 K \cite{Wang_PRL_2012,Sobota_PRL_2012}. Since in such a state the optical phonon cooling is expected to be the most effective channel, the fast-decaying component can be attributed to an electron-optical-phonon scattering process. The corresponding scattering rate $\tau_{ep}^{-1}$ as a function of $T$ ($T\gtrsim$10 K) is shown in Fig. \ref{fig:deltaR}(b), where the linear-in-$T$ behavior is expected for a given Debye temperature of $\Theta_D\sim180$ K \cite{Lai_APL_2014}. The following slow relaxation has a time constant of $\sim$1 ps, in well agreement with previous findings of the electron-phonon scattering time associated with the low energy phonons \cite{Sobota_PRL_2012, Wang_PRL_2012, Hajlaoui_NL_2012, Crepaldi_PRB_2012, Qi_APL_2010, Kumar_PRB_2011, Hsieh_PRL_2011, Onishi_PRB_2015}. We can extract the oscillatory component in $\Delta R/R$, whose Fourier transform (FT) reveals several frequencies [as shown in Fig. \ref{fig:deltaR}(c)], e.g., $\sim$2.2 THz, 4 THz, and 5.2 THz at 293 K. These terahertz oscillations are due to coherent optical phonons, initiated via either coherent Raman scattering \cite{Garrett_PRL_1996} or displacive excitation \cite{Cheng_APL_1991}. The three peaks in Fig. \ref{fig:deltaR}(c) from left to right are attributed to $A_{1g}^1$, $E_g^2$, and $A_{1g}^2$ Raman-active optical phonon modes, respectively \cite{Richter_PSS_1977}. Another $E_g^1$ Raman-active mode can also be observed at low temperatures (see Supplemental Material \cite{Suppl}). Among all the optical phonon modes, $A_{1g}^1$ is the strongest, suggesting that electron-$A_{1g}^1$-optical-phonon coupling dominates the e-p scattering or the momentum scattering time ($\tau_p$) in the excited bulk states within the initial $\sim$500 fs, as also indicated by the Tr-ARPES measurements \cite{Sobota_PRL_2014}. 

In addition, one can unambiguously extract component $(\Delta R/R)_{E_g}$ from $\Delta R/R$ [Fig. \ref{fig:deltaR}(d)], which directly shows the electron-phonon scattering process coupled solely with the $E_g$ phonon modes ($E_g^2$ dominates, see Supplemental Material \cite{Suppl}). Fitting $(\Delta R/R)_{E_g}$ with an exponential decay, we obtain an electron-phonon scattering time of $\tau_{ep}^*\simeq$30 fs, which is nearly temperature independent. Clearly, $\tau_{ep}^*$ is about 10 times faster than $\tau_{ep}$ associated with the dominant $A_{1g}^1$ phonon mode. Our experiment thus evidently demonstrates that scattering events coupled with different phonon modes will decay in distinct timescales.

\begin{figure}
\includegraphics[width=17cm]{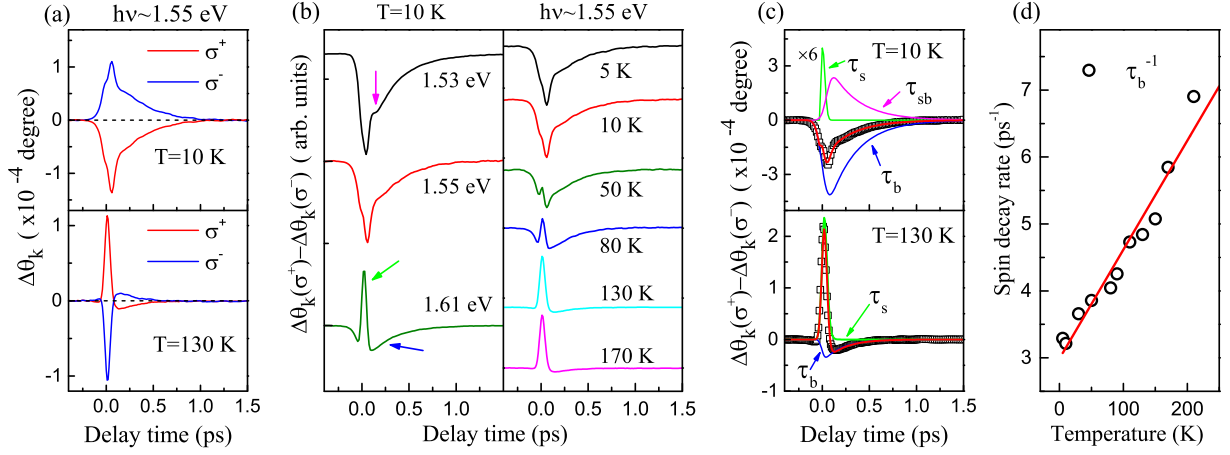}
\caption{\label{fig:TRKR}(a) Time-resolved Kerr rotation, $\Delta\theta_K$, induced by left ($\sigma^+$) and right ($\sigma^-$) circularly polarized light at 10 K and 130 K. (b) Excitation photon energy ($h\nu$) and temperature ($T$) dependence of $[\Delta\theta_K(\sigma^+)-\Delta\theta_K(\sigma^-)]$, where $\Delta\theta_K(\sigma^+)-\Delta\theta_K(\sigma^-)\simeq2\Delta\theta_K(\sigma^+)$. (c) Exponential decay fittings (red lines) for experimental $\Delta\theta_K$ at 10 K and 130 K with $h\nu\sim1.55$ eV (black squares). Green, magenta, and blue lines describe three distinct dynamical processes characterized by $\tau_s$, $\tau_{sb}$, and $\tau_b$, respectively. (d) Spin decay rate $\tau_b^{-1}$ as a function of $T$. The red line is a linear fit.}
\vspace*{-0.2cm}
\end{figure}

Figure \ref{fig:TRKR}(a) shows typical measurements of $\Delta\theta_K$ at low and high temperatures. It can be seen that the circularly polarized light generates a non-equilibrium net spin polarization in Bi$_2$Se$_3$, as the Kerr rotation signals change sign with the helicity of the pump light. This observation is consistent with that reported in Ref. \cite{Hsieh_PRL_2011}, where the spin relaxation time is $\sim$200 fs (limited by time-resolution) for excited electrons in either SSs or bulk states of Bi$_2$Se$_3$ at room temperature. In this work, our TRKR signals evidently reveal novel complex structures. 

Microscopically, generation of a net spin density in opaque materials by circularly polarized light involves direct optical transitions between different energy levels and the accompanying angular momentum transfer (spin selection rules), as in GaAs \cite{Meier_OptOri_1984}. Therefore, in order to understand the nature of the related electronic states, we have carried out detailed measurements on the $h\nu$-dependent and $T$-dependent $\Delta\theta_K$. The results are shown in Figs. \ref{fig:TRKR}(a)-(c). Here, we first notice that the $h\nu$-dependence of $\Delta\theta_K$ resembles the $T$-dependence. This similarity can be understood by the band-gap shrinkage (BGS) with increasing $T$ via electron-phonon interactions commonly seen in most semiconductors \cite{Cardona_RMP_2005}. Due to the BGS effect, the influence on optical transitions by increasing $T$ is analogous to that by increasing $h\nu$.  

From Figs. \ref{fig:TRKR}(a)-(c), we also find that three dynamical processes contribute to $\Delta\theta_K(t)$. An ultrafast transient with positive sign (indicated by green arrows) dominates at large $h\nu$ or high $T$, but almost disappears at small $h\nu$ or low $T$. It lasts for about 100 fs. A relatively slow decay component with negative sign (indicated by blue arrows) is observed at all investigated $h\nu$ and $T$. It is persistent to $\sim$1.2 ps. The third process with positive sign (indicated by magenta arrows) appears at small $h\nu$ and low $T$. It also lasts for about 1.2 ps. Quantitatively, these three processes can be fitted using three exponential decays with time constants $\tau_s$, $\tau_b$, and $\tau_{sb}$, respectively. The fitted results agree well with the experimental data [Fig. \ref{fig:TRKR}(c)]. Similar data have also been obtained in highly $n$-doped samples (see Supplemental Material \cite{Suppl}).  

\begin{figure}
\includegraphics[width=8.5cm]{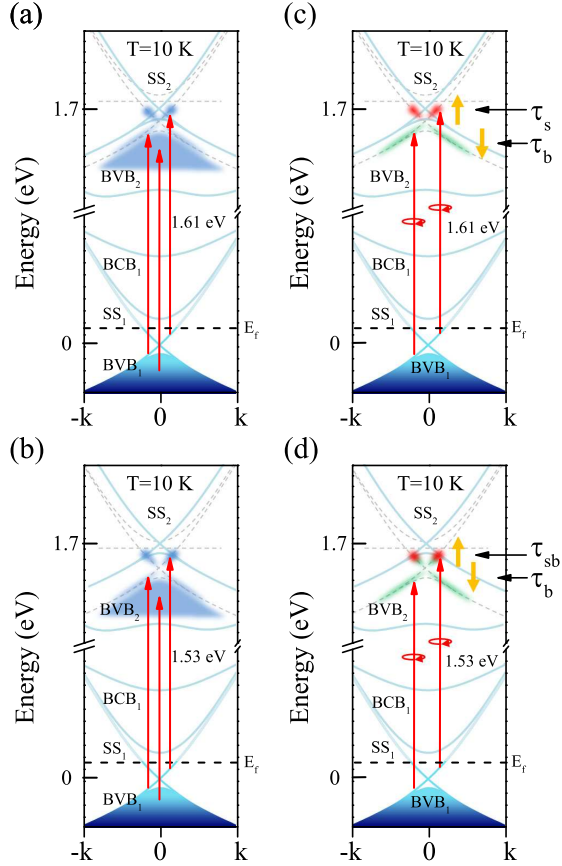}
\caption{\label{fig:TRKR-pic} Schematic illustration of the resonant photo-excitation processes in Bi$_2$Se$_3$ based on the Tr-ARPES results in Refs. \cite{Sobota_PRL_2012, Sobota_JESR_2014}. (a) and (b) show the electrons resonantly populated by the direct optical transitions SS$_1\rightarrow$SS$_2$, SS$_1\rightarrow$BVB$_2$, and BVB$_1\rightarrow$BVB$_2$ (indicated by red arrows) using linearly polarized light at $\sim$1.61 eV and $\sim$1.53 eV, respectively. (c) and (d) show the spin-polarized electrons resonantly excited via SS$_1\rightarrow$SS$_2$ (red shading) and SS$_1\rightarrow$BVB$_2$ (green shading) using circularly polarized light at $\sim$1.61 eV and $\sim$1.53 eV, respectively. Excited electrons generated via SS$_1\rightarrow$SS$_2$ in (a) and (c) are above the top of BVB$_2$, and thus weakly coupled to BVB$_2$. Excited electrons generated via SS$_1\rightarrow$SS$_2$ in (b) and (d) are at the edge of BVB$_2$, where SS$_2$ and BVB$_2$ are strongly coupled.  The spin-polarized electrons generated via SS$_1\rightarrow$SS$_2$ have an opposite polarization direction to that via SS$_1\rightarrow$BVB$_2$ (indicated by up and down yellow arrows). In (d), the spin-polarized electrons are expected to transfer from SS$_2$ into BVB$_2$, and thus exhibit a spin relaxation $\tau_{sb}\simeq\tau_b$, being distinct from $\tau_s$. In all panels, black dashed lines represent the Fermi level $E_f$. The upper dashed lines are guides to the eye for projection of the initial states to $\sim$1.61 eV or $\sim$1.53 eV higher in energy. Because the electrons generated via BVB$_1\rightarrow$BVB$_2$ in (a) and (b) are essentially unpolarized, they are not shown in (c) and (d).}
\vspace*{-0.4cm}
\end{figure}

According to the previous Tr-ARPES studies \cite{Sobota_PRL_2013, Sobota_JESR_2014}, the instantaneous populated states from the optical transitions with $h\nu\sim$1.61 eV should involve the second topological SS (SS$_2$) above the bulk conduction band (BCB$_1$), as shown in Fig. \ref{fig:TRKR-pic}. In this situation, the allowable direct optical transitions are from the first topological SS (SS$_1$) to SS$_2$, from the first bulk valence band (BVB$_1$) to the second high-lying bulk valence band (BVB$_2$), and from SS$_1$ to BVB$_2$ [Fig. \ref{fig:TRKR-pic}(a)]. These transitions are symbolized by SS$_1\rightarrow$SS$_2$, BVB$_1\rightarrow$BVB$_2$, and SS$_1\rightarrow$BVB$_2$, respectively. However, when $h\nu$ or $T$ is relatively small, $h\nu$ becomes insufficient for populating SS$_2$ via SS$_1\rightarrow$SS$_2$, although BVB$_1\rightarrow$BVB$_2$ and SS$_1\rightarrow$BVB$_2$ can still occur. Apparently, if the process characterized by $\tau_s$ is attributed to the relaxation of the spin-polarized electrons in SS$_2$ via SS$_1\rightarrow$SS$_2$, the corresponding TRKR signal should vanish with decreasing $h\nu$ or $T$, in well consistent with our experimental observations [Fig. \ref{fig:TRKR}]. This scenario is also consistent with the theoretical works of Refs. \cite{Hosur_PRB_2011, Lu_PRB_2010}, that is, the spin-polarized electrons can be generated by SS-to-SS interband transitions with particular spin selection rules using circularly polarized light. 

The spin dynamics characterized by $\tau_b$ occurs at all $h\nu$ and $T$ investigated, where both transitions BVB$_1\rightarrow$BVB$_2$ and SS$_1\rightarrow$BVB$_2$ are always allowed [Fig. \ref{fig:TRKR-pic}]. However, the electrons excited by $\sigma^{\pm}$ photons via interband transitions between bulk states should be essentially unpolarized, due to the bulk inversion symmetry \cite{Barriga_arXiv_2015}. Even if for some other reasons, a net spin polarization is generated from the transition between bulk states BVB$_1\rightarrow$BVB$_2$, it is expected to be $h\nu$- and $T$-independent, as the total number of the populated bulk states remains nearly intact in the energy range investigated [Figs. 3(a) and (b)]. But, experimentally, we find that the amplitude of $\tau_b$ process exhibits a strong $h\nu$ and $T$-dependence: it decreases with increasing $h\nu$ and $T$, as shown in Figs. 2(b) and (c). Such dependence is consistent with the population change associated with SS$_1\rightarrow$BVB$_2$. Therefore, one can naturally attribute $\tau_b$ to the spin relaxation of excited electrons in BVB$_2$, as illustrated in Figs. 3(c) and (d). It is worth noting that, due to angular momentum conservation, the spin-polarized electrons in bulk states excited from SSs should have an opposite polarization direction to that in SSs excited from bulk states. This phenomenon is indeed observed in highly $n$-doped samples, where additional transition BCB$_1\rightarrow$SS$_2$ induces a surface net spin polarization in SS$_2$ opposite to the spin state in BVB$_2$ via SS$_1\rightarrow$BVB$_2$ ($\tau_b$ process) (see Supplemental Material \cite{Suppl}).         

The spin dynamics characterized by $\tau_{sb}$ only appears at small $h\nu$ and $T$. It has the same sign as the dynamics characterized by $\tau_s$ (SS$_1\rightarrow$SS$_2$), whereas with a similar relaxation timescale as the dynamics characterized by $\tau_b$ (SS$_1\rightarrow$BVB$_2$). Such observations suggest that this dynamics might be associated with both excited SSs and bulk states, i.e., SS$_2$ (same sign) and BVB$_2$ (similar timescale). In fact, as $h\nu$ or $T$ decreases, the non-equilibrium spin-polarized electrons excited from SS$_1$ to SS$_2$ will be at the edge of BVB$_2$ [Fig. \ref{fig:TRKR-pic}(d)], where strong coupling between SS$_2$ and BVB$_2$ exists \cite{Yazyev_PRL_2010}. The energetic spin-polarized electrons can then efficiently transfer from SS$_2$ to available bulk states in BVB$_2$, while maintaining the polarization direction as in the dynamics characterized by $\tau_s$. This interpretation also leads to $\tau_{sb} \approx\tau_b$, as they both describe the spin relaxation of the spin-polarized electrons in BVB$_2$. It naturally explains the unusual slow rise time ($\sim$150 fs) in the $\tau_{sb}$ process, which is needed for the charge transfer.   

As an intermediate conclusion, our experiments unveil that generation of net spin polarization in three-dimensional TIs using $\sigma^{\pm}$ photons requires the participation of the topological SSs, e.g., SS$_1$ and/or SS$_2$ in Bi$_2$Se$_3$. Remarkably, we find that by properly tuning $h\nu$ or $T$, the spin-polarized electrons excited to unoccupied SSs (e.g., SS$_2$ in Bi$_2$Se$_3$) can have distinct spin relaxation times, depending on whether they are strongly coupled to bulk states or not. Such tunability is essential for the development of future ultrafast spintronic devices \cite{Zutic_RMP_2004}. Therefore, the underlying mechanism behind it requires to be further explored. 

In a TI system such as Bi$_2$Se$_3$, the electron spin precessing around an effective magnetic field with corresponding frequency $\Omega$ during a correlation time $\tau_c$ (average timescale of spin precession along one direction) may experience two different spin relaxation mechanisms. The first is the D'yakonov Perel' (DP) mechanism for SSs with spin-momentum locking in absence of the inversion symmetry \cite{Zutic_RMP_2004}. In this case, the strong spin-orbit coupling (SOC) in SSs induces the effective magnetic field with the corresponding $\Omega$ given by the energy splitting between $E_{\overrightarrow{k}\downarrow}$ and $E_{\overrightarrow{k}\uparrow}$: $\hbar\Omega\simeq2\hbar v_fk$ \cite{Zutic_RMP_2004,Hasan_RMP_2010,Qi_RMP_2011}, where $\vec{k}$ is the electron wave vector, $\hbar$ is the reduced Planck's constant, $v_f$ is the Fermi velocity, and $\uparrow\downarrow$ stand for spins with opposite directions. Taking $\hbar\Omega\sim2E_f\simeq0.2$ eV and $\tau_c\simeq\tau_p\sim50$ fs from infrared transmission measurements \cite{Butch_PRB_2010}, we thus can obtain $\Omega\tau_c\gg1$. The second spin relaxation mechanism is the Elliot-Yafet (EY) mechanism for bulk states with inversion symmetry \cite{Zutic_RMP_2004}. Here, the spin-orbit interaction, altered by phonons, causes the couping of different spin states and leads to the spin flips via momentum scattering. The correlation time $\tau_c$ is then determined by the momentum scattering time $\tau_p$ and the inverse of the frequency $f^{-1}$ of relevant thermal phonons \cite{Zutic_RMP_2004}, both of which give $\tau_c\sim$500 fs, taking $\tau_p\sim\tau_{ep}$ and $f^{-1}\sim f^{-1}(A_{1g}^1)$. In addition, $\hbar\Omega$ can be estimated by the strength of SOC in Bi$_2$Se$_3$, with a typical value of $\hbar\Omega\simeq\lambda_{SO}\sim 0.1$ eV \cite{Qi_RMP_2011, Liu_PRB_2010}. In fact, the coupling strength between bulk states BCB$_j$ and BVB$_j$ ($j=1,2$) can be as large as the energy gap ($\sim0.3$ eV) \cite{Liu_PRB_2010}. $\Omega\tau_c\gg1$ then can also be obtained for the bulk states in EY mechanism. 

The relation $\Omega\tau_c\gg1$ implies that the electron spin precesses many full cycles during $\tau_c$ around the effective magnetic field. The spin polarization decays irreversibly after $\tau_c$ \cite{Zutic_RMP_2004}. Therefore, the spin relaxation time $\tau^{spin}$ is given by $\tau^{spin}\sim\tau_c$, applied to both excited SSs and bulk states in Bi$_2$Se$_3$, which is consistent with recent theoretical works \cite{Burkov_PRL_2010,Liu_PRL_2013,Zhang_PRB_2013}.

Experimentally, we reveal two distinct relaxation timescales for the coherent spins excited in SSs and bulk states. The fast dynamics is characterized by $\tau_s\sim 25$ fs, which is nearly $T$-independent and related to the excited spin-polarized electrons in SS$_2$. Since $\tau_s\sim\tau_{ep}^*$, one can conclude that the electron-$E_g^2$-optical-phonon scattering dominates the scattering process in excited SS, and hence $\tau_s$. This result is in contrast to that in the excited bulk states, where two relatively slow spin relaxation $\tau_b$ and $\tau_{sb}$ associated with BVB$_2$ are found to be $\sim$300 fs at low $T$, similar to $\tau_{ep}$ ($\tau_b \simeq \tau_{sb} \sim \tau_{ep}$) [Figs. 1(b) and 2(d)]. In addition, a linear-in-$T$ dependence is observed for $\tau_b^{-1}$, which exactly follows the $T$ dependence of $\tau_{ep}^{-1}$. These observations thus suggest that the scattering/spin relaxation in excited bulk states is dominated by the $A_{1g}^1$ phonon mode. Since the $E_g^2$ and $A_{1g}^1$ phonon modes have distinct energies, the spin dynamics in topological SSs and bulk states exhibit different behavior. This finding, along with previous works \cite{Sobota_PRL_2014, Zhu_PRL_2011, Zhu_PRL_2012, Costache_PRL_2014}, may inspire further investigation on dynamical properties in topological SSs.        

In summary, we performed ultrafast optical spectroscopy study of spin dynamics in TI Bi$_2$Se$_3$. We unravel that non-equilibrium net spin polarization induced by $\sigma^{\pm}$ photons requires the participation of topological surface states (SSs) in the interband transitions. For the first time, we demonstrate that besides manipulating non-equilibrium electron spins in TI Bi$_2$Se$_3$ by switching circular polarization, one can selectively excite spin-polarized electrons to unoccupied SS$_2$ with two distinct spin relaxation times by only tuning the photon energy or temperature. We reveal that the distinct spin relaxation arise from the scattering in SSs and bulk states which is dominated by $E_g^2$ and $A_{1g}^1$ phonon modes, respectively. Our measurements thus pave a way to manipulate the photoinduced coherent spins in TIs, which may have profound implications in future TI-based ultrafast spintronic devices.       

We would like to thank Hongming Weng and Yan Li for helpful discussions. This work was supported by China 1000-Young Talents Plan, National Natural Science Foundation of China (Grants Nos. 10979021, 11027401, 11174054, 11304338 and 11227902), the Ministry of Science and Technology of China (National Basic Research Program Grant No. 2011CB921800), the “Strategic Priority Research Program (B)” of the Chinese Academy of Sciences (Grant No. XDB04010100) and Helmholtz Association through the Virtual Institute for Topological Insulators (VITI). Z. Jiang acknowledges the support from the U.S. Department of Energy (Grant No. DE-FG02-07ER46451).\\

\begin{center}
\textbf{\Large Supplemental Material}
\end{center}

\section{\label{sec:level11} Sample preparation and experimental setup}

In our experiment, the high quality Bi$_2$Se$_3$ single crystals were grown by the Bridgman method. The samples used in the main text were synthesized by mixing Bi and Se with a non-stoichiometric ratio of 2:3.9 during the growth process. This growing method leads to a carrier density of $n\simeq4\times10^{18}$ cm$^{-3}$, compared with the nominally-stoichiometric mixed samples (Bi:Se=2:3) where $n$ is larger by about one order of magnitude \cite{Wang_PRL_2012}. We have also carried out measurements on such highly-doped Bi$_2$Se$_3$ crystals with carrier density of $n\simeq3\times10^{19}$ cm$^{-3}$, corresponding to Fermi level $E_f$ inside the bulk conduction band (BCB$_1$, as defined in the main text) and $\sim$0.3 eV above the Dirac point. Some detailed experimental results on this type of samples are shown in this Supplemental Material. Prior to optical measurements, Bi$_2$Se$_3$ crystals were cleaved along the \textit{a-b} plane under high vacuum ($\sim10^{-7}$ Torr). Each $\Delta R(t)/R$ or $\Delta\theta_K(t)$ curve in main text can be collected within $\sim$5 minutes. 

\begin{figure}[h]
	\includegraphics[width=12cm]{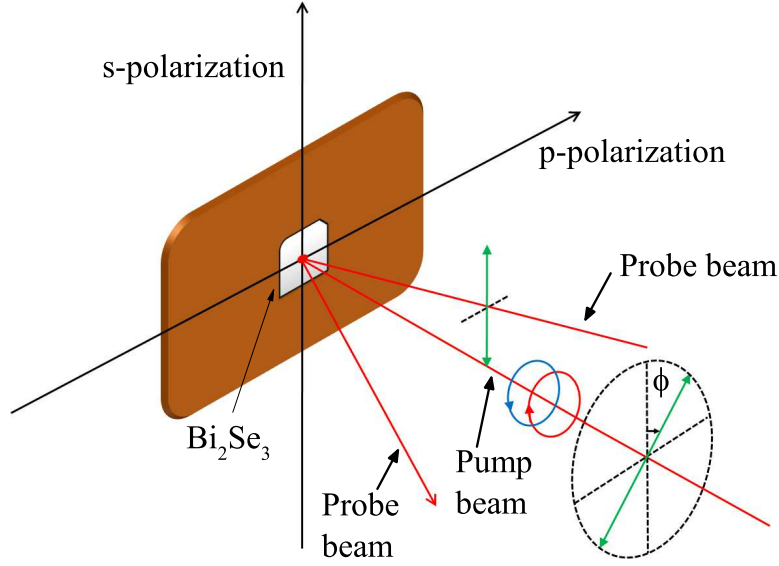}
	\caption{\label{fig:pump-probe} Schematic of the experimental geometry.}
\end{figure}

\begin{figure}[b]
	\includegraphics[width=9cm]{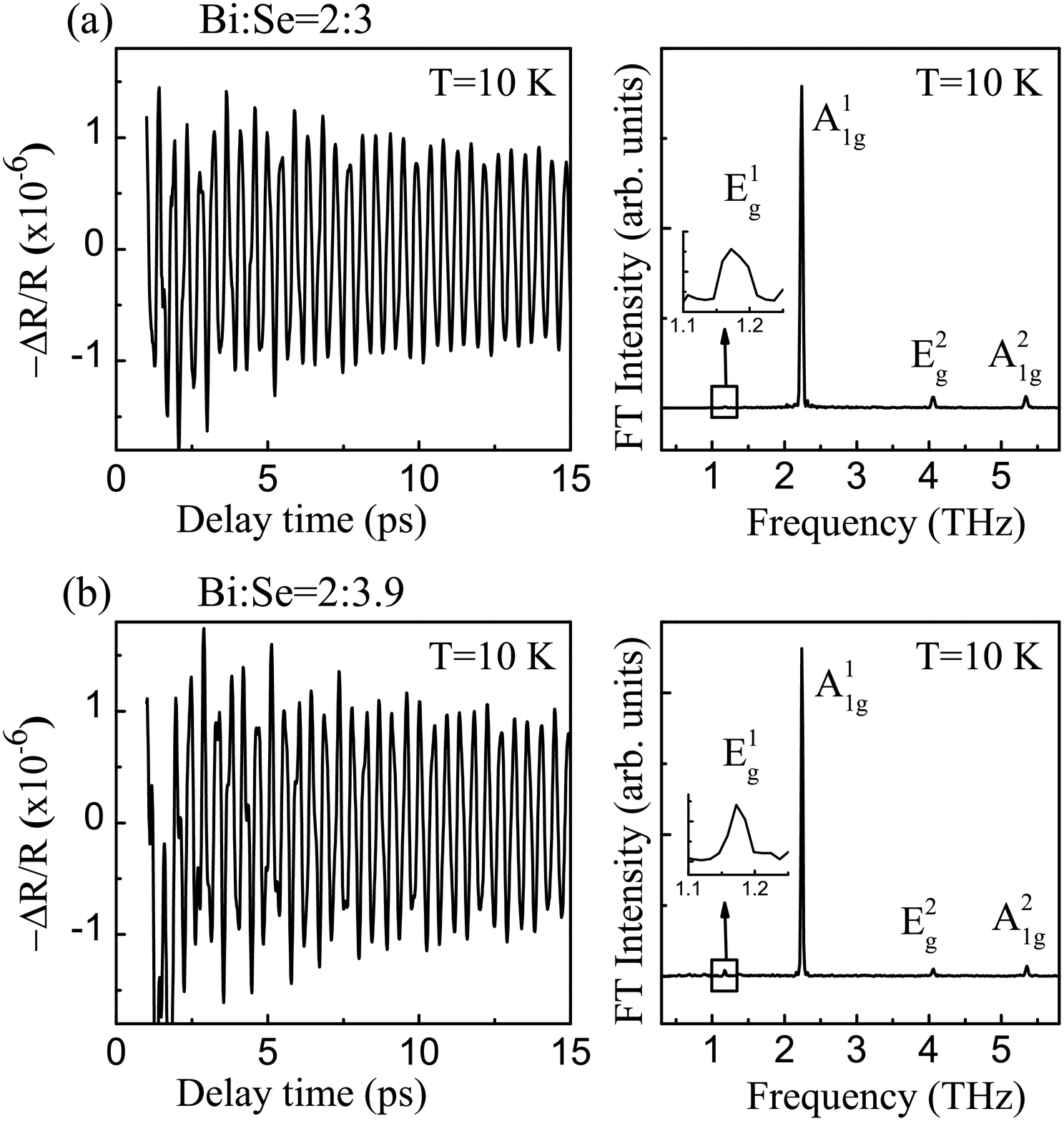}
	\caption{\label{fig:oscillation} Extracted oscillations from $\Delta R/R$ and their corresponding FT spectrum at 10 K for both low- and high-density Bi$_2$Se$_3$ samples. Four Raman-active optical phonon modes are identified.}
	\vspace*{-0.2cm}
\end{figure}

The time-resolved transient reflectivity change, $\Delta R/R$, and the time-resolved Kerr rotation (TRKR), $\Delta\theta_K$, were measured based on a degenerate pump-probe scheme using a Ti:sapphire laser oscillator that produces $\sim$25 fs pulses with a repetition rate of 80 MHz at center wavelengths ranging from 770 nm ($\sim$1.61 eV) to 810 nm ($\sim$1.53 eV) [Fig. \ref{fig:pump-probe}]. Time resolution, $\sim$35 fs, is determined by the FWHM value of the pump-probe cross-correlation profile at the sample position. Typically, we use wavelength 800 nm ($\sim$1.55 eV) for the measurements unless noted otherwise. The pump beam directs along the normal. The probe beam is incident at a $\sim$10 degree angle to the sample normal. In the experiment, the polarization of the probe beam is kept $s-$polarized, while the pump polarization has an angle of $\phi$ to the probe polarization. Except in angle-dependent measurements, the pump and probe beams are cross-polarized, i.e., $\phi=\pi/2$. In TRKR experiments, the polarization rotation is measured using a balanced photo-detector, and the pump is either left ($\sigma^+$) or right ($\sigma^-$) circularly polarized. Pump and probe beams have fluences of 0.3 $\mu$J/cm$^2$ and 0.03 $\mu$J/cm$^2$, respectively. Further experimental details can also be found in our previous works \cite{Qi_PRB_2009, Qi_APL_2010, Qi_PRL_2013}

\section{\label{sec:level12} Additional experimental results}
\subsection{\label{sec:level21} Raman-active optical phonon modes revealed by $\Delta R/R$}

Fourier transform (FT) of the extracted oscillatory component at 10 K reveals four frequencies for high-density (a) and low-density (b) Bi$_2$Se$_3$ samples [Fig. \ref{fig:oscillation}]: $\sim$1.17 THz, 2.2 THz, 4.03 THz, and 5.4 THz. These terahertz oscillations are due to coherent optical phonons, initiated either via coherent Raman scattering \cite{Garrett_PRL_1996} or displacive excitation \cite{Cheng_APL_1991}. The four peaks in FT spectrum from left to right are attributed to $E_g^1$, $A_{1g}^1$, $E_g^2$, and $A_{1g}^2$ Raman-active optical phonon modes, respectively \cite{Richter_PSS_1977}. 

\begin{figure}
	\includegraphics[width=16cm]{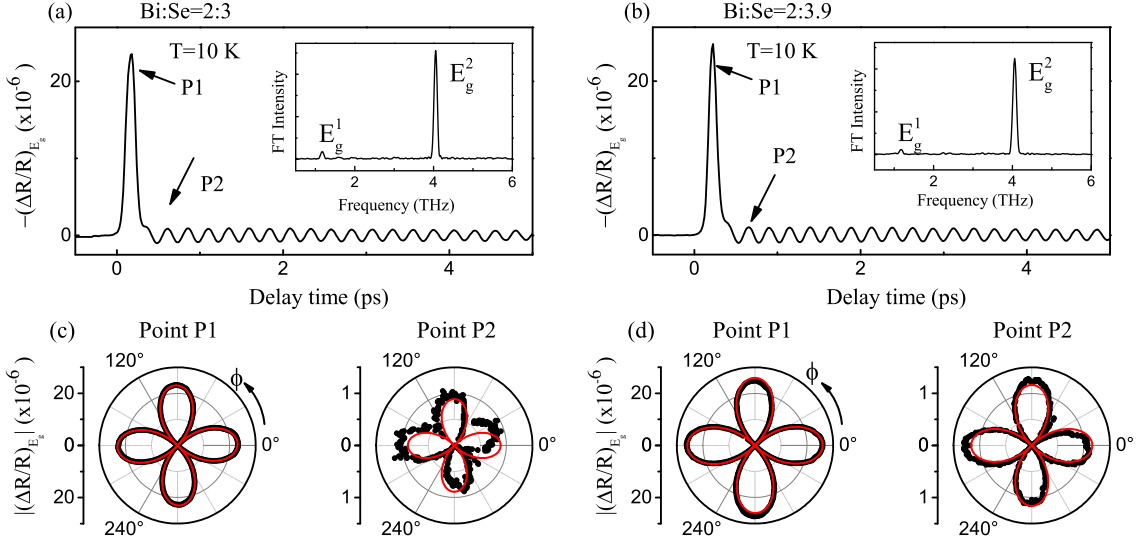}
	\caption{\label{fig:deltaR_Eg} Typical $(\Delta R/R)_{E_g}$ signals associated with $E_g$ phonon modes at 10 K for high-density (a) and low-density (b) Bi$_2$Se$_3$ samples. (c) and (d) illustrate the rotational anisotropy of $(\Delta R/R)_{E_g}$ as a function of the angle $\phi$ at points P1 and P2 in (a) and (b), respectively. Insets to (a) and (b) are FT spectra of the extracted oscillatory component.}
	\vspace*{-0.2cm}
\end{figure}

\subsection{\label{sec:level22} Transient reflectivity $(\Delta R/R)_{E_g}$ associated with $E_g$ phonon modes}

In pump-probe experiment, the transient reflectivity change associated with phonon field $Q$ can be described by a Raman-like process \cite{Garrett_PRL_1996}. Phenomenologically, the phonon field $Q$ excited by the pump light is given by \cite{Garrett_PRL_1996}
\begin{equation}
\frac{d^2Q}{dt^2}+\omega^2Q=\overrightarrow{F}(\overrightarrow{r},t),
\end{equation}
where $\omega$ is the corresponding phonon frequency, and $\vec{F}$ is the driving force. Within coherent Raman scattering mechanism $F=\Sigma_{\mu\nu}(R_{\mu\nu}^RE_{\mu}E_{\nu})/2$, where $E_{\mu}$ and $E_{\nu}$ denote the components of the optical pump field, $R_{\mu\nu}^R\approx\partial\chi_{\mu\nu}/\partial Q$ is the nonlinear Raman-like tensor, and $\chi_{\mu\nu}$ is the linear susceptibility.

\begin{figure}
	\includegraphics[width=12cm]{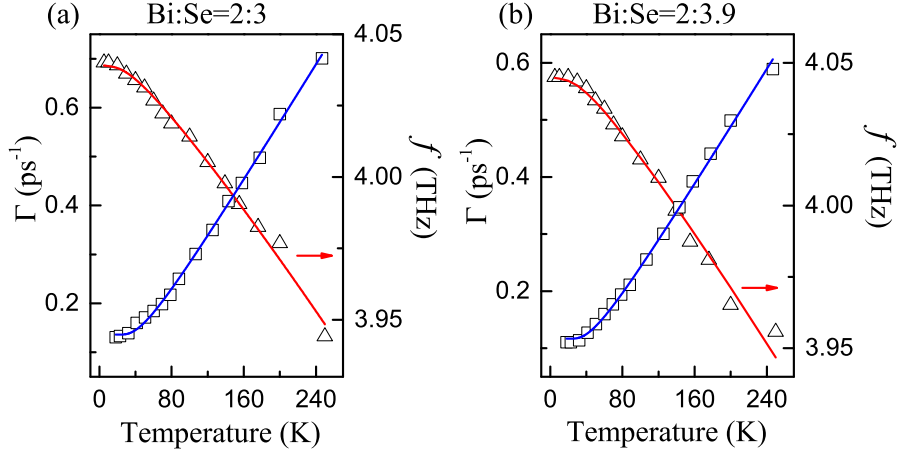}
	\caption{\label{fig:EgVsT} $T$-dependence of the oscillation frequency $f$ and the damping rate $\Gamma$ for the $E_g^2$ phonon mode. The solid curves are fits to the data using the anharmonic decay model, Eqs. (2) and (3).}
	\vspace*{-0.2cm}
\end{figure}

Therefore, based on coherent Raman scattering selection rules derived from the Raman tensors of Bi$_2$Se$_3$ \cite{Richter_PSS_1977}: $A_{1g}$ $\begin{pmatrix} a & 0 & 0 \\ 0 & a & 0 \\ 0 & 0 & b \end{pmatrix}$ and $E_{g}$ $\begin{pmatrix} 0 & -c & -d \\ -c & 0 & 0 \\ -d & 0 & 0 \end{pmatrix}$ $\begin{pmatrix} c & 0 & 0 \\ 0 & -c & d \\ 0 & d & 0 \end{pmatrix}$, intensity of the $A_{1g}$ mode is independent of angle $\phi$, while the $E_g$ mode behaves as cos$2\phi$ \cite{Garrett_PRL_1996}. $\phi$ is the angle between pump and probe polarizations, as described above. The transient reflectivity associated with $E_g$ phonon mode thus can be written as $[\Delta R(t,\phi)/R]_{E_g}=[\Delta R(t)/R]_{E_g}$cos$2\phi$. As a consequence, $[\Delta R(t)/R]_{E_g}$ can be experimentally obtained, e.g., by the difference between measurements of $\Delta R(t)/R$ at $\phi=\pi/2$ and at $\phi=0$. Figure \ref{fig:deltaR_Eg} shows the typical experimental results of $(\Delta R/R)_{E_g}$, as well as the cos$2\phi$ dependence for two selected points P1 and P2 at different time delays. FT spectrum of the oscillatory component shows that only $E_g^1$ and $E_g^2$ phonon modes are present, and the $E_g^2$ phonon mode dominates the electron-phonon scattering process in $(\Delta R/R)_{E_g}$. The peak corresponding to the $E_g^1$ phonon mode in the FT spectra could not be observed when the temperature is above $\sim$70 K. Therefore, below we will only focus on the temperature dependence of the oscillation frequency $f$ and damping rate $\Gamma$ for the dominant $E_g^2$ mode. The results are shown in Fig. \ref{fig:EgVsT}, where the experimental $f$ and $\Gamma$ are obtained by fitting the oscillatory component in $(\Delta R/R)_{E_g}$ using $Ae^{-\Gamma t}sin(2\pi ft+\beta)$ \cite{Qi_PRL_2013}. These results are consistent with the previous work using conventional Raman spectroscopy \cite{Kim_APL_2012} and the temperature dependence can be well explained by the anharmonic phonon decay model (solid lines in Fig. \ref{fig:EgVsT}) \cite{Kim_APL_2012},
\begin{align}
\omega(T)=\omega_0+\Delta\omega^{(1)}(T)+A_1[1+2n(\omega_0/2)], 
\\
\Gamma(T)=A_2[1+2n(\omega_0/2)],
\label{eq:Phonon-fit}
\end{align} 
where $\omega=2\pi f$, $n(\omega)=[e^{\hbar\omega/k_BT}-1]^{-1}$, $\omega_0$ is the bare harmonic frequency, and the shift $\Delta\omega^{(1)}$ from thermal expansion is given by $\Delta\omega^{(1)}(T)=\omega_0[e^{-\gamma\int\limits_{0}^{T}(\alpha_c+2\alpha_a)dT^{\prime}}-1]$. The Gr\"uneisen parameter $\gamma$ and the thermal expansion coefficients $\alpha_i$ $(i=a, c)$ are obtained from Ref. \cite{Chen_APL_2011}.

\begin{figure}
	\includegraphics[width=15cm]{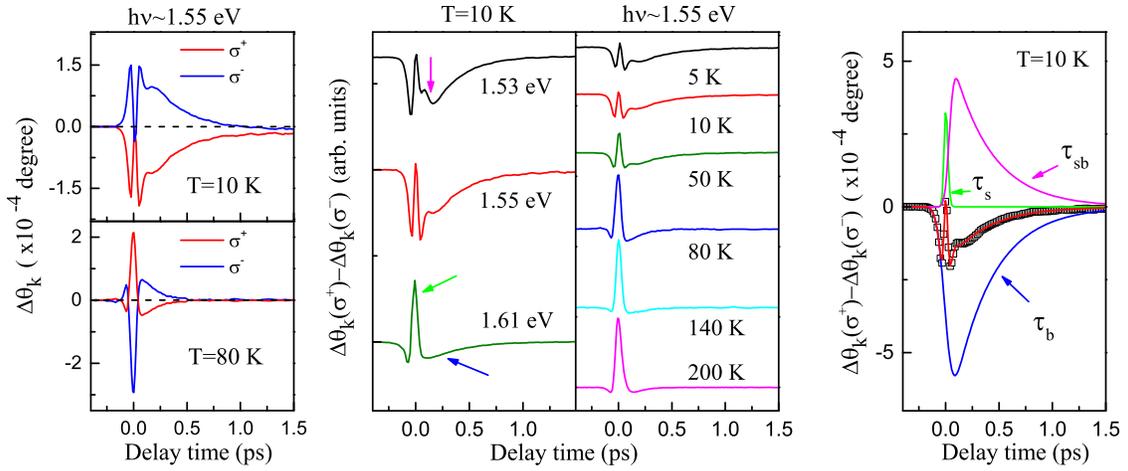}
	\caption{\label{fig:TRKR-hd} TRKR results on highly-doped Bi$_2$Se$_3$ samples. Left panel: time-resolved Kerr rotation, $\Delta\theta_K$, induced by left ($\sigma^+$) and right ($\sigma^-$) circularly polarized light at 10 K and 80 K. Middle panel: excitation photon energy ($h\nu$) and temperature ($T$) dependence of $[\Delta\theta_K(\sigma^+)-\Delta\theta_K(\sigma^-)]$, where $\Delta\theta_K(\sigma^+)-\Delta\theta_K(\sigma^-)\simeq2\Delta\theta_K(\sigma^+)$. Right panel: exponential decay fittings (solid red line) for $\Delta\theta_K$ at 10 K with $h\nu=1.55$ eV. Green, magenta, and blue lines describe three distinctive dynamical processes characterized by $\tau_s$, $\tau_{sb}$, and $\tau_b$, respectively.}
	\vspace*{-0.4cm}
\end{figure}

\subsection{\label{sec:level23} TRKR results on highly-doped Bi$_2$Se$_3$ samples}

\begin{figure}
	\includegraphics[width=8cm]{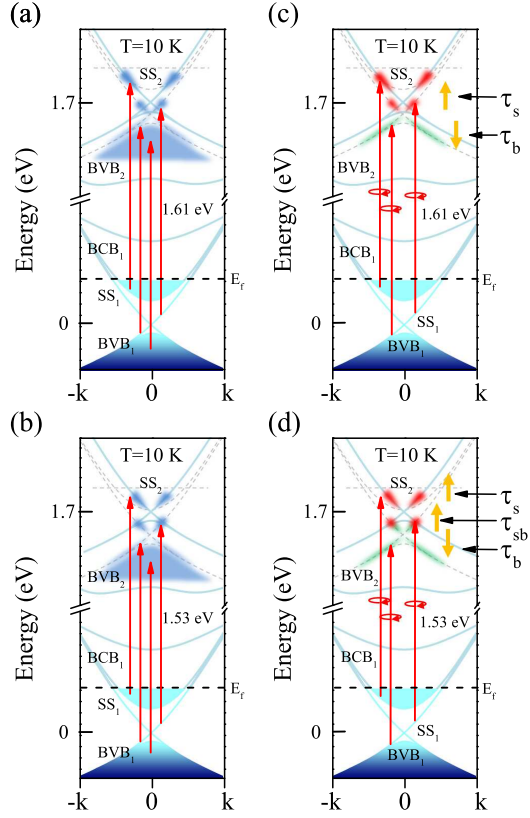}
	\caption{\label{fig:TRKR-pic-hd} Schematic illustration of the resonant photo-excitation processes in highly-doped Bi$_2$Se$_3$ samples. (a) and (b) show electrons resonantly populated by the direct optical transitions SS$_1\rightarrow$SS$_2$, SS$_1\rightarrow$BVB$_2$, BVB$_1\rightarrow$BVB$_2$, and BCB$_1\rightarrow$SS$_2$ (indicated by red arrows) at $\sim$1.61 eV and $\sim$1.53 eV, respectively. (c) and (d) show the spin-polarized electrons resonantly excited via SS$_1\rightarrow$SS$_2$ (red shading), BCB$_1\rightarrow$SS$_2$ (red shading), and SS$_1\rightarrow$BVB$_2$ (green shading) using circularly polarized light at $\sim$1.61 eV and $\sim$1.53 eV, respectively. The excited electrons generated via SS$_1\rightarrow$SS$_2$ and BCB$_1\rightarrow$SS$_2$ in (a) and (c) are above the top of BVB$_2$, and thus only weakly coupled to BVB$_2$. The excited electrons generated via SS$_1\rightarrow$SS$_2$ in (b) and (d) are at the edge of BVB$_2$, where SS$_2$ and BVB$_2$ are strongly coupled. The spin polarized electrons generated via SS$_1\rightarrow$SS$_2$ and BCB$_1\rightarrow$SS$_2$ have an opposite polarization direction to that via SS$_1\rightarrow$BVB$_2$ (indicated by up and down yellow arrows, and red and green shadings). In (d), the spin polarized electrons generated via SS$_1\rightarrow$SS$_2$ will transfer from SS$_2$ into BVB$_2$, and show a spin relaxation $\tau_{sb}\simeq\tau_b$, being distinct from $\tau_s$ in excited SS$_2$. In all panels, black dashed lines represent the Fermi level $E_f$. The upper dashed lines are guides to the eye for projection of the initial states to $\sim$1.61 eV or $\sim$1.53 eV higher in energy, respectively. Because the electrons generated via BVB$_1\rightarrow$BVB$_2$ are essentially unpolarized, they are not shown in (c) and (d).}
\end{figure}

Figure \ref{fig:TRKR-hd} demonstrates that the three spin dynamics characterized by $\tau_s$, $\tau_{sb}$, and $\tau_b$ (indicated by green, magenta, and blue arrows, respectively), discussed in our main text, can also be clearly seen in the highly-doped Bi$_2$Se$_3$ crystals. Comparing with the low-density samples, the main difference here is that the ultrafast transient component characterized by $\tau_s$ remains even at the smallest $h\nu$ and lowest $T$ investigated. This phenomenon is due to an additional optical transition BCB$_1\rightarrow$SS$_2$ involved photo-excitation process [Fig. S6], as the Fermi level $E_f$ moves into BCB$_1$ \cite{Sobota_PRL_2013}. Such an optical transition is allowed at the smallest $h\nu$ and lowest $T$ that we could reach. The spin-polarized electrons can be generated in SS$_2$ via BCB$_1\rightarrow$SS$_2$ (see discussion in the main text). Their polarization direction is opposite to that in BVB$_2$ via SS$_1\rightarrow$BVB$_2$, while their spin relaxation time are the same as that in SS$_2$ via SS$_1\rightarrow$SS$_2$ [Fig. \ref{fig:TRKR-pic-hd}]. Therefore, the spin dynamics characterized by $\tau_s$ arising from the excited SS$_2$ can always survive under our experimental conditions. 

\subsection{\label{sec:level24} Influence of aging effect on time-resolved Kerr rotation measurements}

\begin{figure}
	\includegraphics[width=12cm]{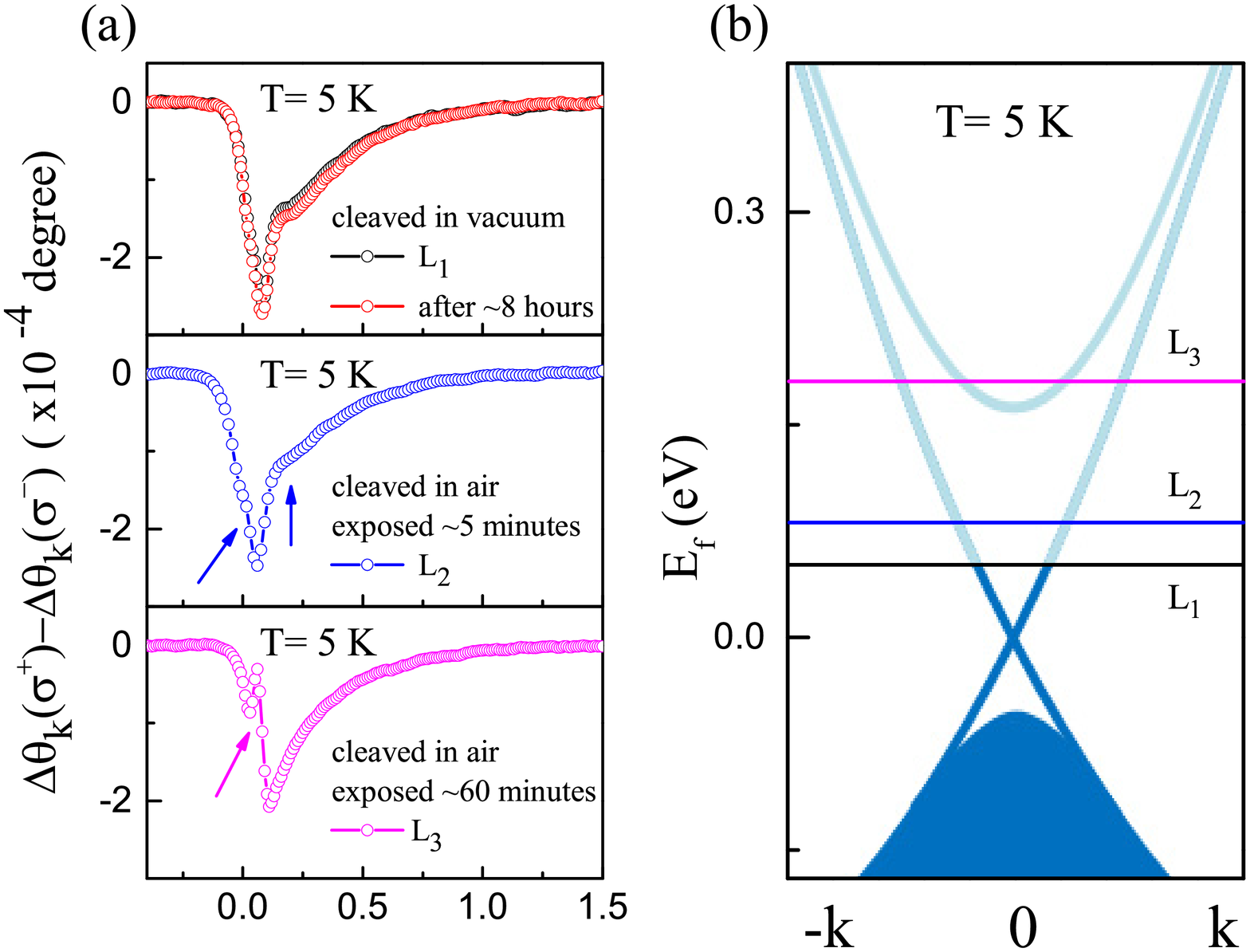}
	\caption{\label{fig:TRKR-aging} (a) TRKR results obtained on the low-density Bi$_2$Se$_3$ samples at $T$=5 K and $h\nu\sim$1.55 eV under different aging conditions. (b) Schematic of the estimated Fermi level (indicated by L$_1$, L$_2$, and L$_3$) under different aging conditions. Black (L$_1$), red, blue (L$_2$) and pink (L$_3$) curves in (a) represent four aging conditions: (1) cleaved in vacuum and measured immediately after cleavage, (2) cleaved in vaccum and measured after $\sim$8 hours, (3) cleaved in air, $\sim$5 minutes air exposure, and then measured in vacuum, (4) cleaved in air, $\sim$60 minutes air exposure, and then measured in vacuum, respectively. }
\end{figure}

Although the Bi$_2$Se$_3$ crystals are cleaved and measured in high vacuum, we still need to evaluate whether aging effect or surface doping \cite{Benia_PRL_2011,King_PRL_2011} will affect our experimental data and interpretation. Towards this end, we have measured the low-density crystals under four different aging conditions: (1) cleaved in vacuum and measured immediately after cleavage; (2) cleaved in vaccum and measured after $\sim$8 hours; (3) cleaved in air, $\sim$5 minutes air exposure, and then measured in vacuum; (4) cleaved in air, $\sim$60 minutes air exposure, and then measured in vacuum. In these conditions, the aging effect or surface doping becomes stronger from (1) to (4). The experimental results are summarized in Fig. \ref{fig:TRKR-aging}.

It can be clearly seen from Fig. \ref{fig:TRKR-aging} that $\Delta\theta_K$ shows little difference between samples in conditions (1) and (2). Therefore, when interpreting our experimental data, we can safely neglect the aging effect or surface doping within our experimental resolution, as all our measurements were conducted within $\sim$3 hours ($\lesssim$30 minutes for $h\nu$-dependent data, and $\lesssim$3 hours for $T$-dependent data) after \textit{in situ} cleavage. In addition, surface doing effect was indeed observed in samples under aging conditions (3) and (4). In Fig. \ref{fig:TRKR-aging} (a), we notice that $\Delta\theta_K$ for samples in condition(3) with $\sim$5 minutes air exposure already displays some changes (indicated by the blue arrows), compared to the samples cleaved in vacuum. These changes correspond to appearance of ultrafast transient characterized by $\tau_s$ and decrease of the amplitude of the $\tau_{sb}$ process. When the samples were exposed to air for more time, e.g., $\sim$60 minutes, these changes become more pronounced (see bottom panel of [Fig. \ref{fig:TRKR-aging} (a)]). 

As is already known, stronger aging effect or surface doping leads to a higher Fermi energy $E_f$, as illustrated in Fig. \ref{fig:TRKR-aging} (b). In our experiment, we find that under aging condition (4), the ultrafast transient (indicated by pink arrow in Fig. \ref{fig:TRKR-aging} (a)) in $\Delta\theta_K$ signal is still clearly present at 5 K. This observation is similar with the low-$T$ results obtained on the high-density Bi$_2$Se$_3$ samples, which have a Fermi level sitting inside BCB$_1$. Our interpretation in main text suggests that: (a) the increase of $E_f$ would result in the increased amplitude for the $\tau_s$ process and decreased amplitude for the $\tau_{sb}$ process, (b) $E_f$ sitting inside BCB$_1$ would lead to the strong presence of the $\tau_s$ process at lowest $T$ that we can reach for $h\nu\sim$1.55 eV. Thus, our interpretation is well consistent with the observations shown in Fig. \ref{fig:TRKR-aging} (a). 

Therefore, our experimental data on aging effect provide additional support to the interpretation in main text on the photoinduced coherent spin dynamics.  

\subsection{\label{sec:level25} Fitting methods}

All fittings to the measured signals $\Delta R/R$, $(\Delta R/R)_{E_g}$, and $\Delta\theta_K$ (excluding the oscillatory components) are accomplished by \cite{Hilton_PRL_2002}

\begin{equation}
Signals=\sum_{j=1}^{m}A_je^{-t/\tau_j}\otimes G_j(t),
\end{equation}
where $\tau_j$ is the decay time constant, and $m$ ($\leq3$) depends on how many exponential decays are involved. $G_j(t)$ is the normalized cross correlation of $g_0(t)$ and $g_j(t)$, given by
\begin{equation}
G_j(t)=g_0(t)\otimes g_j(t).
\end{equation}
Here, $g_0(t)$ represents the probe pulse, which has a Gaussian shape with a FWHM of 35 fs. $g_j(t)$ is also a Gaussian, whose FWHM is determined by fitting the rise time of each pump-induced physical process characterized by $\tau_j$.


\begin{thebibliography}{text}
\bibitem{Hasan_RMP_2010} M. Z. Hasan and C. L. Kane, Rev. Mod. Phys. {\bf 82}, 3045 (2010).
\bibitem{Qi_RMP_2011} X.-L. Qi and S.-C. Zhang, Rev. Mod. Phys. {\bf 83}, 1057 (2011).
\bibitem{Tse_PRL_2010} W. -K Tse and A. H. MacDonald, Phys. Rev. Lett. 105, 057401 
\bibitem{Aguilar_PRL_2012} R. Valdes Aguilar et al., Phys. Rev. Lett. {\bf 108}, 087403 (2012).
\bibitem {Inoue_PRL_2010} J. I. Inoue and A. Tanaka, Phys. Rev. Lett. {\bf 105}, 017401 (2010).
\bibitem{Lindner_NatPhys_2011} N. H. Lindner, G. Refael, and V. Galitski, Nature Phys. {\bf 7}, 490 (2011).
\bibitem{Wang_Science_2013} Y. H. Wang, H. Steinberg, P. Jarillo-Herrero and N. Gedik Science {\bf 342}, 453 (2013).
\bibitem{McIver_NatNano_2011} J. W. McIver, D. Hsieh, H. Steinberg, P. Jarillo-Herrero, and N. Gedik, Nat. Nanotechnol. {\bf 7}, 96 (2011).
\bibitem{Lu_PRB_2010} H. Z. Lu, W. Y. Shan, W. Yao, Q. Niu, and S. Q. Shen, Phys. Rev. B {\bf 81}, 115407 (2010).
\bibitem{Hosur_PRB_2011} P. Hosur, Phys. Rev. B {\bf 83}, 035309 (2011).
\bibitem{Junck_PRB_2013}A. Junck, G. Refael, and F. von Oppen, Phys. Rev. B {\bf 88}, 075144 (2013).
\bibitem{Kastl_NatComm_2015} C. Kastl, C. Karnetzky, H. Karl, and A. W. Holleitner, Nat. Comm. {\bf 6}, 6617 (2015).
\bibitem{Pesin_NatMat_2012} D. Pesin and A. H. Macdonald, Nat. Mater. {\bf 11}, 409 (2012).
\bibitem{Sobota_PRL_2012} J. A. Sobota et al., Phys. Rev. Lett. {\bf 108}, 117403 (2012).
\bibitem{Wang_PRL_2012} Y. H. Wang et al., Phys. Rev. Lett. {\bf 109}, 127401 (2012).
\bibitem{Hajlaoui_NL_2012} M. Hajlaoui et al., Nano Lett. {\bf 12}, 3532 (2012).
\bibitem{Crepaldi_PRB_2012} A. Crepaldi et al., Phys. Rev. B {\bf 86}, 205133 (2012).
\bibitem{Sobota_PRL_2013} J. A. Sobota et al., Phys. Rev. Lett. {\bf 111}, 136802 (2013).
\bibitem{Sobota_JESR_2014} J. A. Sobota et al., J. Electron Spectrosc. Relat. Phenom. {\bf 149}, 295 (2014).
\bibitem{Sobota_PRL_2014} J. A. Sobota et al., Phys. Rev. Lett. {\bf 113}, 157401 (2014).
\bibitem{Cacho_PRL_2015} C. Cacho et al., Phys. Rev. Lett. {\bf 114}, 097401 (2015).
\bibitem{Qi_APL_2010} J. Qi et al., Appl. Phys. Lett. {\bf 97}, 182102 (2010).
\bibitem{Kumar_PRB_2011} N. Kumar et al., Phys. Rev. B {\bf 83}, 235306 (2011).
\bibitem{Hsieh_PRL_2011} D. Hsieh et al., Phys. Rev. Lett. {\bf 107}, 077401 (2011).
\bibitem{Chen_APL_2012} H.-J. Chen et al., Appl. Phys. Lett. {\bf 101}, 121912 (2012).
\bibitem{Glinka_APL_2013} Y. D. Glinka et al., Appl. Phys. Lett. {\bf 103}, 151903 (2013).
\bibitem{Luo_NL_2013} C. W. Luo et al., Nano Lett. {\bf 13}, 5797 (2013).
\bibitem{Lai_APL_2014} Y. Lai, H. Chen, K. Wu, and J. Liu, Appl. Phys. Lett. {\bf 105}, 232110 (2014).
\bibitem{Cheng_APL_2014} L. Cheng et al., Appl. Phys. Lett. {\bf 104}, 211906 (2014).
\bibitem{Sim_PRB_2014} S. Sim et al., Phys. Rev. B 89, 165137 (2014).
\bibitem{Aguilar_APL_2015} R. V. Aguilar et al., Appl. Phys. Lett. {\bf 106}, 011901 (2015).
\bibitem{Onishi_PRB_2015} Y. Onishi et al., Phys. Rev. B {\bf 91}, 085306 (2015).
\bibitem{Butch_PRB_2010} N. P. Butch et al., Phys. Rev. B {\bf 81}, 241301 (2010).
\bibitem{Zutic_RMP_2004} I. Zutic, J. Fabian, and S. D. Sarma, Rev. Mod. Phys. {\bf 76}, 323 (2004); M. W. Wu, J. H. Jiang, and M. Q. Weng, Phys. Rep. {\bf 493}, 61236 (2010); M. I. D'yakonov, ed., \textit{Spin Physics in Semiconductors} (Springer-Verlag Berlin Heidelberg, 2008).
\bibitem{Barriga_arXiv_2015} J. Sanchez-Barriga et al., arXiv:1505.02742. 
\bibitem{Boschini_arXiv_2015} F. Boschini et al., arXiv:1506.02692.
\bibitem{Suppl} See Supplemental Material, which includes Refs.\cite{Qi_PRB_2009,Qi_PRL_2013,Kim_APL_2012,Chen_APL_2011,Benia_PRL_2011,King_PRL_2011,Hilton_PRL_2002}.
\bibitem{Garrett_PRL_1996} G. A. Garrett, T. F. Albrecht, J. F. Whitaker, and R. Merlin, Phys. Rev. Lett. {\bf 77}, 3661 (1996); R. Merlin, Solid State Comm., {\bf 102}, 207 (1997).
\bibitem{Cheng_APL_1991} T. K. Cheng et al., Appl. Phys. Lett. {\bf 59}, 1923 (1991).
\bibitem{Richter_PSS_1977} W. Richter, H. Kohler, and C. R. Becker, phys. stat. sol. (b) {\bf 6}, 619 (1977)
\bibitem{Meier_OptOri_1984} F. Meier and B. P. Zakharchenya, eds., Optical Orientation (Elsevier, Amsterdam,1984).
\bibitem{Cardona_RMP_2005} M. Cardona and M. L. W. Thewalt, Rev. Mod. Phys. {\bf 77}, 1173 (2005).
\bibitem{Yazyev_PRL_2010} O. V. Yazyev, J. E. Moore, and S. G. Louie, Phys. Rev. Lett. {\bf 105}, 266806 (2010).
\bibitem{Liu_PRB_2010} C. X. Liu et al., Phys. Rev. B {\bf 82}, 045122 (2010).
\bibitem{Burkov_PRL_2010} A. A. Burkov and D. G. Hawthorn, Phys. Rev. Lett. {\bf 105}, 066802 (2010).
\bibitem{Liu_PRL_2013} X. Liu, and J. Sinova, Phys. Rev. Lett. {\bf 111}, 166801 (2013).
\bibitem{Zhang_PRB_2013} P. Zhang and M. W. Wu, Phys. Rev. B {\bf 87}, 085319 (2013).
\bibitem{Zhu_PRL_2011} X. Zhu et al., Phys. Rev. Lett. {\bf 107}, 186102 (2011).
\bibitem{Zhu_PRL_2012} X. Zhu et al., Phys. Rev. Lett. {\bf 108}, 185501 (2012).
\bibitem{Costache_PRL_2014} M. V. Costache et al., Phys. Rev. Lett. {\bf 112}, 086601 (2014).
\bibitem{Qi_PRB_2009} J. Qi et al., Phys. Rev. B {\bf 79}, 085304 (2009).
\bibitem{Qi_PRL_2013} J. Qi et al., Phys. Rev. Lett. {\bf 111}, 057402 (2013).
\bibitem{Kim_APL_2012} Y. Kim et al., Appl. Phys. Lett. 100, 071907 (2012).
\bibitem{Chen_APL_2011} X. Chen et al., Appl. Phys. Lett. 99, 261912 (2011).
\bibitem{Benia_PRL_2011} H. M. Benia, C. Lin, K. Kern, and C. R. Ast, Phys. Rev. Lett. {\bf 107}, 177602 (2011).
\bibitem{King_PRL_2011} P. D. C. King et al., Phys. Rev. Lett. {\bf 107}, 096802 (2011).
\bibitem{Hilton_PRL_2002} D. J. Hilton and C. L. Tang, Phys. Rev. Lett. {\bf 89}, 146601 (2002).


\end{thebibliography}
\end{document}